\documentclass[twocolumn,showpacs,preprintnumbers,amsmath,amssymb]{revtex4}
\usepackage{graphicx}% Include figure files
\usepackage{dcolumn}% Align table columns on decimal point
\usepackage{bm}% bold math
\usepackage[T1]{fontenc}
\usepackage{ulem} %Pour barrer du texte.
\begin{document}

\preprint{AIP - Appl. Phys. Lett.}

%\title{Structural transition in LaVO$_3$/SrVO$_3$ superlattices and their influence on the transport properties}% Force line breaks with \\
\title{Structural transition in LaVO$_3$/SrVO$_3$ superlattices and its influence on transport properties}

\author{A. David}%
\author{R. Fr\'esard}
\author{Ph. Boullay}
\author{W. Prellier}
\author{U. L\"uders}
 \email{Ulrike.Luders@ensicaen.fr}

\affiliation{%
CRISMAT, UMR CNRS-ENSICAEN 6508; \\
6 boulevard Mar\'echal Juin, 14050 Caen cedex 4, France
}%

\author{P.-E. Janolin}%
\affiliation{%
Laboratoire Structures, Propri\'et\'es et Mod\'elisation des Solides, UMR CNRS-\'Ecole Centrale Paris; \\
92295 Ch\^atenay-Malabry Cedex, France
}%
\date{\today}

\begin{abstract}

Measurements of the resistive properties and the lattice parameters of a (LaVO$_3$)[6 unit cells]/(SrVO$_3$)[1 unit cell] superlattice between 10K and room temperature are presented. A low temperature metallic phase compatible with a Fermi liquid behavior is evidenced. It disappears in the vicinity of a structural transition from a monoclinic to tetragonal phase, in which disorder seems to strongly influence the transport. Our results will enrich the understanding of the electronic properties of complex heterostructures.

\end{abstract}

\maketitle

The interfaces between perovskite oxides such as SrTiO$_3$ and LaAlO$_3$ show a number of peculiar and interesting phenomena: two-dimensional (2D) electron gases or liquids\cite{Oht04,Thi06,Bas08,Hot07}, superconductivity\cite{Rey07}, and even possibly magnetism\cite{Bri07} at low temperatures. Apart from potential applications in post-Si electronics\cite{Man10}, the exact origin of these phenomena is still under investigation, as the interface reconstruction phenomena between such perovskites are not well understood\cite{Nak06}. To create a comparable framework avoiding highly sensitive zones as interfaces as active regions, we introduce geometrically confined doped (GCD) systems and apply the concept to LaVO$_3$/SrVO$_3$ superlattices \cite{She09}. In this particular case, one unit cell (uc) of SrVO$_3$ (SVO) is introduced between insulating LaVO$_3$ (LVO) layers to create conducting zones with 2D character. The substitution of a LaO subplane by a SrO one in the perovskite structure leads to a mixed valence in the adjacent VO$_2$ subplanes with partially localized electrons due to the reduced bandwidth in the doped regions, as indicated by room-temperature magnetism\cite{lud09}. Thus, these systems show signs of a 2D behavior of the conduction electrons, and the effects are more robust against disorder compared to the above mentioned oxide interfaces.

On the other hand, LaVO$_3$ is a material with intriguing properties. It shows strong orbital fluctuations at high temperature \cite{Ray07} and a complex structural and magnetic transition at around 140K \cite{Fuj05}. The influence of this transition on the charge carriers confined to the doped regions has to be elucidated. For this reason, we have investigated the low temperature structure of the superlattice (SL) LVO[6uc]/SVO[1uc], and the correlation between the resistive and structural properties of this superlattice is presented in this Letter.

The sample was prepared by Pulsed Laser Deposition on SrTiO$_3$ (001)-oriented substrates, details are described elsewhere \cite{She07}. The SL consists of 30 repetitions of the bilayer (total thickness 82nm). The deposited number of layers was verified by the analysis of SL satellite peaks in the X-ray diffraction (XRD) pattern (not shown here), the separation in angle of which is directly related to the thickness of the bilayers, and transmission electron microscopy \cite{Bou11}. The in-plane resistance was measured in the four-point mode with surface contacts in a Physical Properties Measurement System by Quantum Design between 300K and 10K. Temperature-dependent x-ray diffraction was carried out between 295K and 12K on a high-precision diffractometer using Cu K$_{\beta}$ wavelength (1.3922\AA) issued from a 18kW rotating anode.

\begin{figure}
\includegraphics[width = 0.5\textwidth]{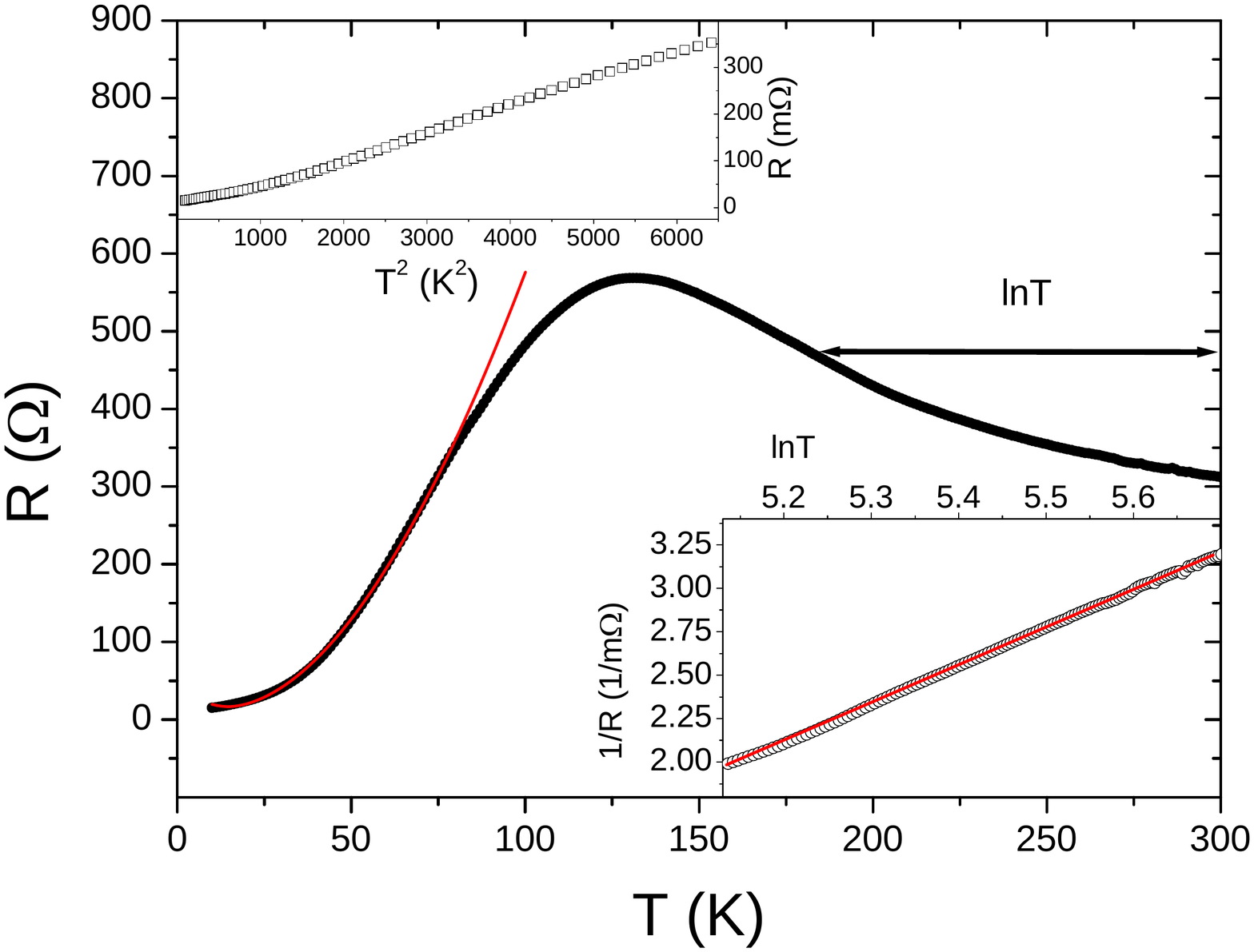}% Here is how to import EPS art
\caption{(color online) Resistance vs temperature of the LaVO$_3$[6uc]/SrVO$_3$[1uc] superlattice and a power-law fit of the low temperature data (red line). The upper inset shows the low temperature resistance data as a function of T$^2$. The lower inset shows the conductivity vs lnT on the temperature range indicated by the arrow in the main graph and a linear fit of the data. }
\label{Figure1}
\end{figure}

The resistance data \textit{vs} temperature of the LVO[6uc]/SVO[1uc] SL is shown in Fig. \ref{Figure1}. Roughly, three different regions can be identified: (i) a temperature dependence with negative slope dR/dT between room temperature and 185K, (ii) a broad transition region with a maximum in the resistance at 130K, and (iii) a low temperature metallic phase down to 10K. Resistivity was not calculated based on this data, as the penetration depth of the current in the SL is not known due to the insulating character of the LVO layers. However, we estimate that the resistivity at 10K is at least 100$\mu\Omega$cm (if the total SL thickness is taken into account) and could reach down to 1$\mu\Omega$cm (if the thickness of the conduction channel corresponds to only one monolayer of doped LVO). The experimental data below 80K can be fitted with a power law, with the best fit shown in Fig. \ref{Figure1}. The temperature exponent of this fit is 1.81(3). As the available temperature range for the fit is relatively small, the data could also be fitted with a T$^2$ dependence. But, since the resistance data as a function of T$^2$ is not strictly linear (see upper inset of Fig. \ref{Figure1}), we are not able to determine the temperature exponent unambiguously. Nevertheless, this metallic phase is noteworthy, as the low resistivity is indicative of a high charge mobility, and as it is neither observed in the solid solution, nor in comparable systems with a ultrathin SrVO$_3$ layer\cite{Hot07}. Also, theoretical description of the LVO/SVO interface \cite{Jac08} and STO/SVO superlattices \cite{Par10} fails to predict a metallic phase. 

At high temperatures, although the slope dR/dT is negative, the resistance of the sample stays low with a weak temperature dependence, pointing towards a metallic phase with a loss of charge carrier coherence rather than an insulating phase. Indeed, while activated transport or variable range hopping theories fail, the high temperature region can be fitted with a lnT behavior (see lower inset in Fig. \ref{Figure1}), which can be interpreted as weak localization in 2D \cite{Vol92}. Magnetoresistance measurements in this region show a positive magnetoresistance (not shown here). This, in addition, is indicative of the relevance of the electron-electron interactions as the origin of this weak localization of the charge carriers \cite{Alt80}. 

Since this resistive transition has not been observed in bulk, we believe that its origin could give information on the physics of these superlattices with geometrically confined doping. A maximum in the resistance versus temperature observed in strongly correlated systems is often associated with magnetic transitions, where ferromagnetic ordering of the spins results in a reduction of the resistance due to double exchange coupling, as for example in the case of mixed valence manganites\cite{Rao96}. In our case, the absence of a transition to a ferromagnetic phase in this temperature range suggests that such a mechanism is not at the origin of this resistive transition.

Interestingly, the change of coherence at the resistive transition indicates a change in disorder of the system, which may have its origin in the structural transition of bulk LVO at around 140K, where the high temperature orthorhombic structure (space group Pnma) changes to a low temperature monoclinic structure (space group P2$_1$/a)\cite{Bor93}. The microstructural investigation of the superlattice by electron microscopy confirms that the room temperature structure of the LaVO$_3$ in the SL is comparable to the bulk structure\cite{Bou11}. To verify the presence of a structural phase transition reminiscent of the bulk one in the SL, its structure was characterized between 10K and room temperature. 

\begin{figure}
\includegraphics[width = 0.5\textwidth]{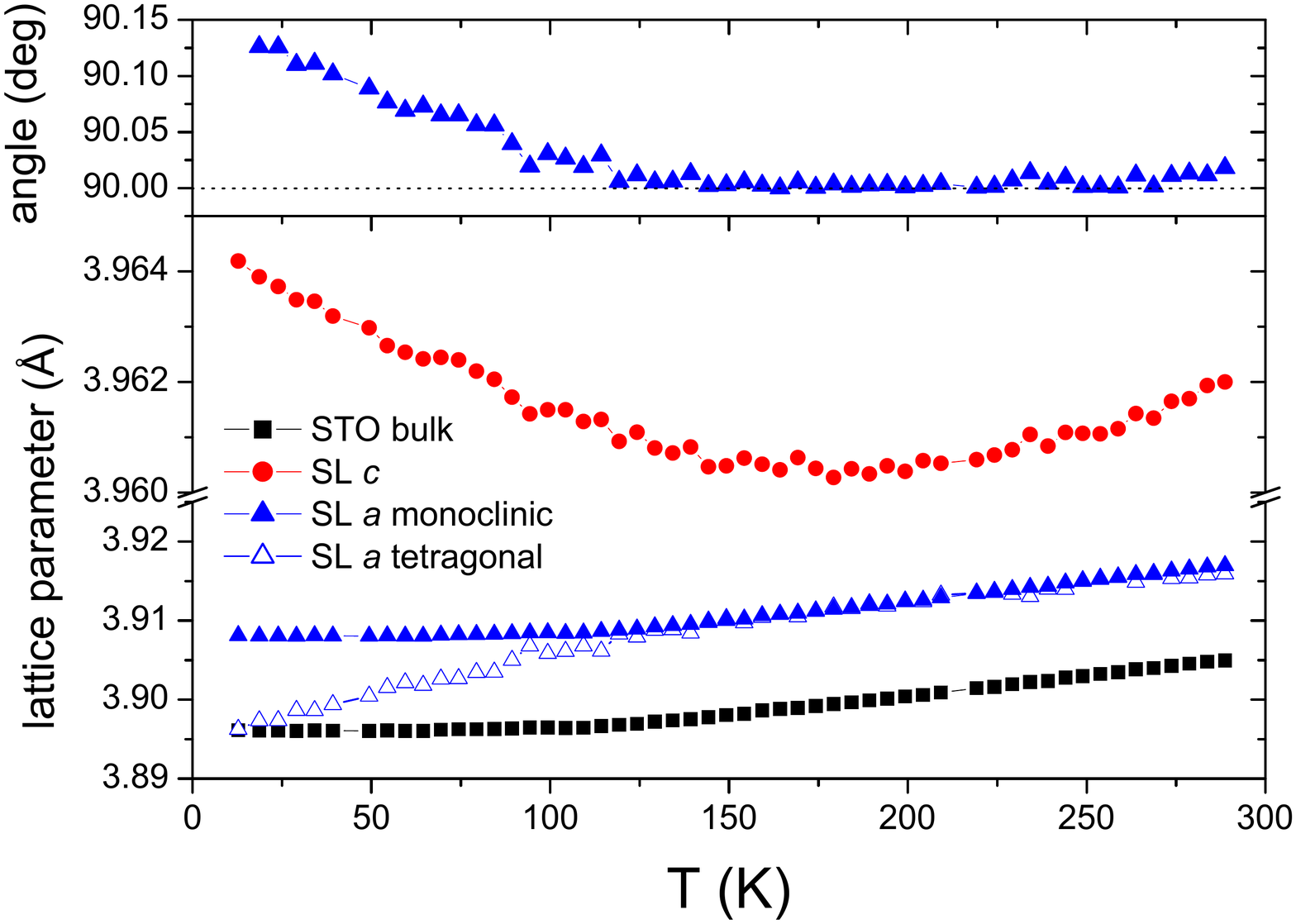}% Here is how to import EPS art
\caption{(color online) Evolution with temperature of the out-of-plane (filled red dots) and in-plane (triangles) lattice parameters of the superlattice. The black squares represent the lattice parameter of the substrate (SrTiO$_3$). The solid (empty) triangles represent the temperature evolution of the in-plane lattice constant of the superlattice within the monoclinic (tetragonal) symmetry and under the assumption of a constant density of dislocation. Top panel: Temperature evolution of the monoclinic angle.}
\label{Figure2}
\end{figure}

XRD measurements at room temperature show that the SL is monocrystalline and its $\approx$ 4\AA\ pseudo-cubic (pc) unit cell is oriented such as (001)$^{pc}_{SL}$//(001)$_{STO}$ and [100]$^{pc}_{SL}$ and [010]$^{pc}_{SL}$ are parallel to [100]$_{STO}$ and [010]$_{STO}$, respectively. The pseudo-cubic out-of-plane and in-plane lattice parameters of the SL have been determined from several (00l)$_{l=1-5}$ and (204) reflections at room temperature. Let us underline that the lattice parameters have to be calculated from the available peaks within the hypothesis of a crystal system. The temperature-dependence of the lattice parameters reported in Figure \ref{Figure2} illustrates two hypothesis: a tetragonal one, where the $c$ axis is constrained to be perpendicular to the $a-b$ plane (open triangles) and a monolinic one where it is allowed to tilt towards the $a-b$ plane (solid triangles). As can be seen from Fig.\ref{Figure2}, from 295K to 130K the in-plane lattice parameter of the SL $a_{SL}$ exhibits an evolution parallel to the one of the substrate $a_{STO}$ for both hypothesis. Such an evolution has been reported for other mono-oriented perovskite films \cite{Jan07,Jan07b,Choi04} and is due to the constant density of dislocations that accommodate partially the lattice mismatch between the substrate and the SL. The substrate nevertheless imposes the evolution of its lattice parameters with temperature onto $a_{SL}$\cite{Jan09}. At 130K a phase transition from a tetragonal high-temperature symmetry to a low-temperature monoclinic one takes place. The tetragonal crystal system is not suited below 130K to describe the SL as $a_{SL}$, calculated within this system, would tend toward $a_{STO}$ (open triangles in Fig.\ref{Figure2}), indicating dislocations vanishing. This is highly improbable not only considering the reported similar evolutions \cite{Jan07,Jan07b,Choi04} but also because if such vanishing was to occur, it would have taken place at higher temperature. To account for this fact, in the monoclinic hypothesis the temperature evolution of $a_{SL}$ was therefore constrained to follow the evolution of $a_{STO}$ as a result of the constant number of dislocations. The temperature evolution of the position of the (204) reflection is thus accommodated for by the introduction of a monoclinic angle between the c-axis and the $a-b$ plane. At 10K, the monoclinic angle results in a value of 90.12$^{\circ}$ (see top panel of Fig.\ref{Figure2}), consistent with the value observed in bulk LaVO$_3$ \cite{Bor93}, and reaches 90$^{\circ}$ at 130K and above, confirming the tetragonal symmetry of the SL at high temperature.

The resulting picture of the crystal structures of the SL and the transition between them is therefore comparable to bulk LVO, in agreement with electron microscopy studies\cite{Bou11}. The transition temperature and the value of the monoclinic angle are indeed comparable. Thus, the intergrowth with SrVO$_3$ and the clamping to the STO substrate results in a strain state of the LaVO$_3$ layers modest enough to allow for the bulk crystal system. 

The transition from coherent to incoherent transport observed in the resistance measurements is accompanied by a structural transition from a monoclinic structure to a metrically tetragonal one. As the monoclinic distortion of the SL in the low temperature phase is very small, the expected changes of the overlap of the orbitals and therefore the band structure at the Fermi level are marginal. Indeed, an order-disorder transition is more likely to be at the origin of the change in resistive behavior. Studying the bulk transition of LaVO$_3$, the origin of the higher order in the monoclinic structure is not straightforward to unravel. One of the most striking differences between the monoclinic and the orthorhombic phase of bulk LaVO$_3$ is the number of equivalent sets of V sites: in the orthorhombic phase, all V sites are equivalent, while in the monoclinic phase, two sets of inequivalent V sites form which are ordered along the b axis of the monoclinic cell \cite{Bor93}. As the structure of the low temperature SL seems to be comparable to the bulk structure, such an ordering of the inequivalent V sites is possible, therefore explaining the gain in coherence of the charge carriers in the low temperature phase. 

Summarizing, both in the structure as well in the electronic properties a transition was observed in a LaVO$_3$[6uc]/SrVO$_3$[1uc] superlattice. The structural transition is comparable to the transition in LaVO$_3$ bulk, while the related electronic transition is more puzzling: the coherent, metallic-like transport in the monoclinic phase turns into an incoherent, weakly localized transport in the tetragonal phase. The origin of this transition may be the ordering of inequivalent V sites in the monoclinic phase providing for an enhanced disorder above the structural transition to the tetragonal phase and therefore explaining the transition to incoherent transport. The presented study underlines the rich physics and illustrates the possibility of emerging phases in geometrically confined doped Mott insulators.

A.D., R.F., P.B., W.P. and U.L. gratefully acknowledge the R\'egion
Basse-Normandie and the Minist\`ere de la Recherche
for financial support.

\end{document}